# Navigating Independence: A Survey of Visually Impaired People's Experiences and Needs

Banafshe Marziyeh Bamdad,[1, 2] Manuel Günther,[1] and Alireza Darvishy[2]
[1]Department of Informatics, University of Zurich, Switzerland
[2]Institute of Computer Science, Zurich University of Applied Sciences, Switzerland
bamdad@ifi.uzh.ch

## Abstract

Independent navigation in unfamiliar environments remains a major challenge for blind and visually impaired individuals, despite the availability of assistive technologies. This paper presents the results of a fully accessible online survey investigating navigation experiences, challenges, and technology preferences among people with visual impairments worldwide. The survey was distributed through individuals and organizations supporting visually impaired communities. Our results indicate that smartphone-based applications are the most used digital navigation aids, while a substantial proportion of participants report not using any assistive navigation technology due to cost, accessibility, or usability barriers. Participants reported persistent difficulties in obstacle detection, wayfinding, and navigation in complex environments. Despite a widespread focus on smartphone-based solutions, they expressed a clear preference for wearable and hands-free systems, highlighting a gap between current technology use and user needs. The findings provide a user-centered overview of navigation needs and offer insights into the design and evaluation of future assistive navigation systems.

## Introduction

Independent navigation in unfamiliar environments is a fundamental requirement for autonomy, safety, and social participation. For blind and visually impaired (BVI) individuals, navigating everyday spaces such as streets, public buildings, or transportation hubs often involves significant challenges related to orientation, obstacle avoidance, and wayfinding. Although a range of assistive technologies has been developed, their adoption and effectiveness in real-world settings remain limited.
Limited use of assistive technologies should not be interpreted as a lack of need, but rather as a mismatch between user needs and the accessibility, usability, and practicality of available solutions. As illustrated in Figure 1, while many users rely on handheld solutions such as smartphones in practice, a larger proportion expresses a preference for wearable, hands-free technologies.

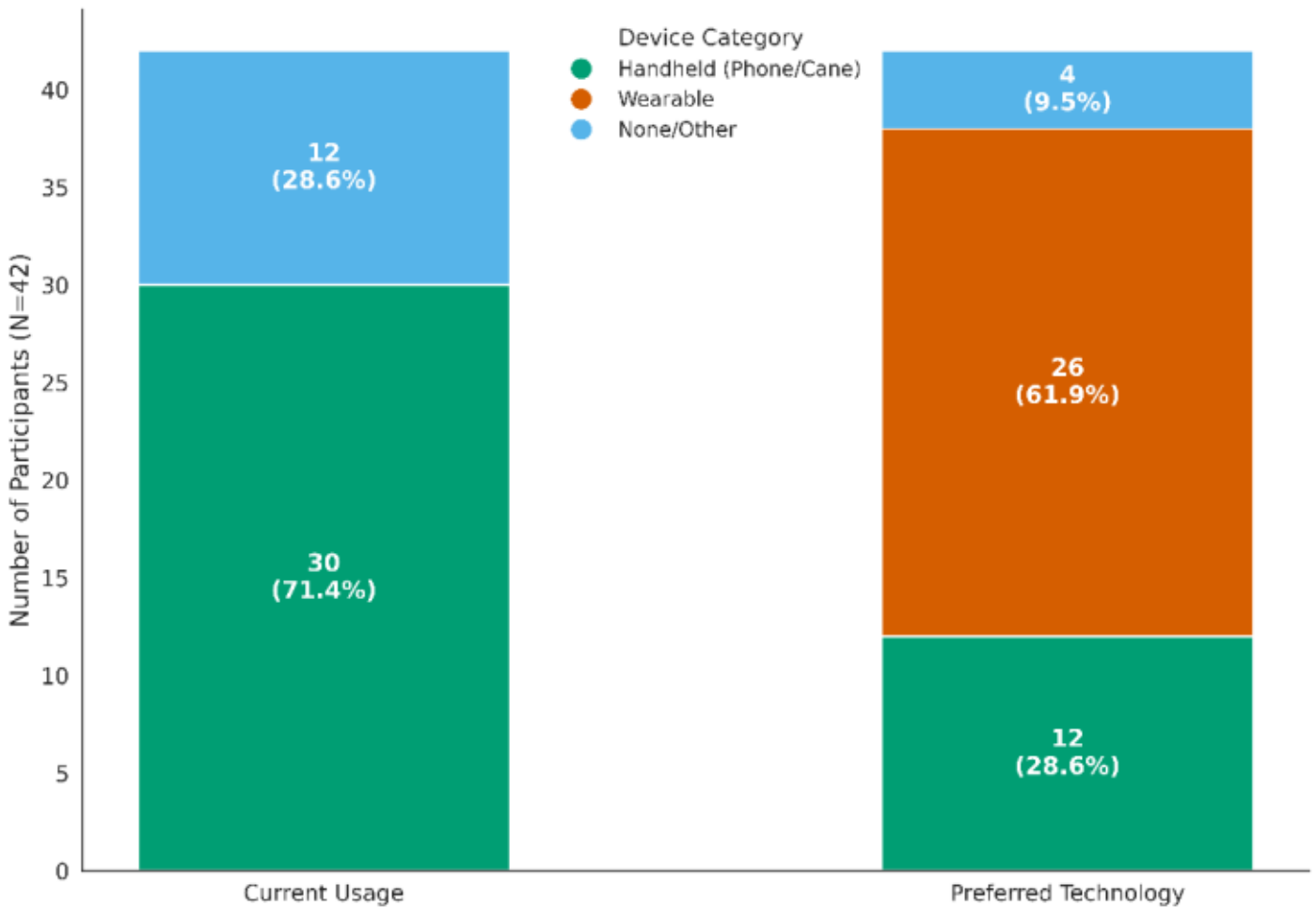

*Figure 1 Comparison of current assistive technology use and preferred device categories, illustrating the gap between commonly used solutions and user preferences.*

What users rely on in practice does not necessarily reflect what they would prefer under ideal conditions. Factors such as cost, ease of use, physical burden, environmental conditions, and regional context strongly influence whether a system is usable in everyday situations. These considerations highlight the importance of grounding assistive navigation research in empirical user experience.
To address this, this paper presents the results of a survey of visually impaired individuals worldwide, capturing navigation experiences, challenges, and preferences related to assistive technologies. The questionnaire included closed- and open-ended questions covering background, technology use, challenges, and navigation needs, and was distributed globally through organizations and individuals supporting visually impaired communities. This work provides a user-centered overview of navigation experiences and needs, complementing existing research that often focuses on technical development. The findings aim to inform and contextualize future research on assistive navigation systems.
The remainder of this paper is organized as follows. First, we review related work, followed by the survey methodology. Afterward, the results of the survey are presented and interpreted, followed by a discussion of limitations, and a conclusion.

## Related Work

Research on navigation for blind and visually impaired individuals has explored a wide range of environments and methodological approaches, including outdoor and indoor settings, as well as studies spanning both contexts. Prior work has employed surveys, interviews, and observational studies to investigate navigation challenges, user needs, and interaction strategies.
A substantial body of work has focused on outdoor navigation, particularly in urban and transportation-related settings [1–7]. These studies highlight challenges related to orientation, environmental awareness, and infrastructure accessibility, and emphasize the role of auditory information and spatial understanding in supporting independent mobility.
Indoor navigation has also received considerable attention, especially in unfamiliar or structured environments such as public buildings and campuses. Prior work has identified key challenges related to wayfinding, obstacle awareness, and the availability of relevant environmental information, and has emphasized the importance of user-centered design in defining effective navigation support [8–10].
Several studies have further examined navigation behavior across different contexts and the use of assistive technologies in everyday scenarios, highlighting the influence of situational factors, user preferences, and interaction strategies [11, 12]. In parallel, a large body of work has proposed and evaluated assistive navigation systems, including wearable, vision-based, and multimodal approaches, often in controlled or system-specific settings [13, 14].
Although these studies provide valuable insights, they are often constrained by specific environments, geographic regions, or particular technologies. In contrast, the present work adopts a user-centered perspective through a globally-distributed and accessible questionnaire, aiming to capture navigation experiences, challenges, and preferences reported by visually impaired individuals across diverse real-world contexts.

## Survey Methodology

This study investigates the experiences, challenges, and preferences of visually impaired individuals in unfamiliar environments through a fully accessible online questionnaire. Accessibility was a central design requirement, and the questionnaire was tested and refined with input from a visually impaired user, supported by manual evaluation using the Web Accessibility Evaluation Tools (WAVE).[1] A pilot study involving visually impaired and sighted participants was conducted to identify usability issues.
The questionnaire consisted of 17 closed-ended questions (12 single-choice and 5 multiple-selection) and two open-ended questions covering personal background, navigation experiences, and preferences. Conditional branching was applied such that follow-up questions were shown only to participants who reported using digital assistive navigation technologies. Eligibility was based on self-identification as blind or visually impaired, verified through early questions on vision status, onset, and severity. The survey aimed to include participants from diverse demographic and geographic backgrounds. Participation was voluntary and based on implied consent, and no personally identifiable information was collected.
Survey responses were cleaned by removing duplicate and incomplete submissions. All questions were required, resulting in a dataset without missing values. Responses were analyzed using quantitative methods for closed-ended questions and inductive thematic analysis [15] for open-ended responses. Given the exploratory nature of the study and the relatively small sample size, the analysis focuses on descriptive statistics. Inferential or multivariate analyses were not performed, as they would not provide statistically reliable conclusions for the present dataset. For key proportions, 95% confidence intervals were computed using the Wilson score interval [16]. The survey questionnaire and supporting documentation are available in the project repository.[2]

[1] https://wave.webaim.org

# Results

This section presents the results of the survey, including participant demographics, experiences with assistive navigation technologies, and stated preferences.

## Participant Background

Participants represented diverse geographic regions and demographic groups, with the largest proportion from Western Europe and North America, followed by Africa and Eastern Europe, as shown in Table 1. The sample included a broad age distribution and a balanced representation of acquired and congenital visual impairment, with most participants reporting severe impairment or total blindness.

*Table 1 Demographics and visual impairment characteristics of 42 participants. Values are reported as number of participants n (%).*

| | | |
|---|---|---|
| **Region** | Africa | 11 (26.2%) |
| | Asia-Pacific | 4 (9.5%) |
| | Eastern Europe | 9 (21.4%) |
| | Latin America & Caribbean | 1 (2.4%) |
| | Western Europe & North America | 17 (40.5%) |
| **Age Group** | Under 15 | 0 (0.00%) |
| | 15-30 | 9 (21.4%) |
| | 30-45 | 14 (33.3%) |
| | 45-60 | 14 (33.3%) |
| | Above 60 | 5 (11.9%) |
| **Gender** | Male | 25 (59.5%) |
| | Female | 16 (38.1%) |
| | Non-binary | 1 (2.4%) |
| **Severity** | Mild | 4 (9.5%) |
| | Moderate | 6 (14.3%) |
| | Severe | 17 (40.5%) |
| | Total blindness | 15 (35.7%) |
| **Onset** | Acquired | 24 (57.1%) |
| | Congenital | 18 (42.9%) |

## Experience with Assistive Navigation Technologies

To understand how these demographic characteristics relate to real-world navigation behavior, we next examine participants' use of assistive technologies. Table 2 presents these results across five categories: technology type, usage settings, overall satisfaction, feedback adequacy and responsiveness (time to react). Smartphone-based applications were the most reported assistive technologies, although a substantial proportion of participants reported not using any digital navigation tools. Nonuse of digital assistive navigation technologies varied across levels of visual impairment. Participants with total blindness and moderate impairment showed higher rates of nonuse (approximately one third), whereas no nonuse was observed among participants with mild impairment. This suggests that individuals with more severe vision loss face greater barriers to adopting existing technologies. Among users, usage was primarily oriented toward outdoor or mixed environments. Satisfaction levels varied, and many participants reported insufficient feedback, particularly in environments with dense obstacles. Perceived responsiveness was also inconsistent, with many users indicating limited time to react to obstacles, highlighting potential safety concerns.
Despite the use of these technologies, participants reported frequent challenges related to route finding, landmark identification, and obstacle detection. Head-level obstacles were reported most often, followed by hazards at chest, ground, and leg levels, reflecting the multi-level nature of real-world navigation risks, as presented in Table 3.

[2] https://github.com/banafshebamdad/PhD-Survey-Visually-Impaired

*Table 2 Experiences with digital assistive navigation of the 32 technology users; reported as n (%), and 95% confidence interval (CI).*

|  |  |  |  |
|---|---|---|---|
| Technology | Smart cane | 2 (4.8%) | [1.3, 15.8] |
| | Smartphone apps | 28 (66.7%) | [51.6, 79.0] |
| | Smartwatch apps | 0 (0.0%) | [0.0, 8.4] |
| | Other technologies | 2 (4.8%) | [1.3, 15.8] |
| | None | 10 (23.8%) | [13.5, 38.5] |
| Setting | Indoor only | 2 (6.3%) | [1.7, 20.1] |
| | Outdoor only | 16 (50.0%) | [33.6, 66.4] |
| | Both in- & outdoor | 14 (43.8%) | [28.2, 60.7] |
| Satisfaction | Very satisfied | 3 (9.4%) | [3.2, 24.2] |
| | Satisfied | 15 (46.9%) | [30.9, 63.6] |
| | Neutral | 4 (12.5%) | [5.0, 28.1] |
| | Unsatisfied | 5 (15.6%) | [6.9, 31.8] |
| | Very unsatisfied | 3 (9.4%) | [3.2, 24.2] |
| | Uncertain | 2 (6.3%) | [1.7, 20.1] |
| Adequacy | Sufficient for navigation | 6 (18.8%) | [8.9, 35.3] |
| | Insufficient indoors | 4 (12.5%) | [5.0, 28.1] |
| | Insufficient outdoors | 8 (25.0%) | [13.3, 42.1] |
| | Insufficient with obstacles | 10 (31.3%) | [18.0, 48.6] |
| | Insufficient in noise | 3 (9.9%) | [3.3, 24.2] |
| | Other limitations | 1 (3.1%) | [0.6, 15.7] |
| Responsiveness | Always sufficient | 2 (6.3%) | [1.7, 20.1] |
| | Usually sufficient | 11 (34.4%) | [20.1, 52.6] |
| | Sometimes sufficient | 12 (37.5%) | [22.9, 54.7] |
| | Rarely sufficient | 4 (12.5%) | [5.0, 28.1] |
| | Never sufficient | 2 (6.3%) | [1.7, 20.1] |
| | Uncertain | 1 (3.1%) | [0.6, 15.7] |

## Preferences and Expectations

In response to these challenges, clear patterns emerged regarding user expectations for assistive technologies. These were examined across three dimensions: device form, feedback modality, and desired system characteristics. Table 4 summarizes the relationship between severity of visual impairment, technology nonuse, and device preferences, with percentages calculated within each severity group. Across all levels of visual impairment, wearable devices were consistently favored, particularly among individuals with severe impairment or total blindness. While handheld solutions were also selected, they were more commonly chosen by participants with severe impairment or total blindness, although they remained less preferred overall. A similar trend was observed across age groups, with wearable devices favored in all categories. As illustrated in Figure 1, a clear discrepancy exists between current usage and preferred device categories: although handheld solutions, particularly smartphones, are most commonly used in practice, a larger proportion of participants expressed a preference for wearable, hands-free technologies.

*Table 3 Navigation challenges and collision risk by obstacle height; reported as n (%), 95% CI; multiple selections allowed.*

|  |  |  |  |
|---|---|---|---|
| Challenges | Finding indoor landmarks | 22 (52.9%) | [37.7, 66.6] |
| | Finding outdoor landmarks | 24 (57.1%) | [42.2, 70.9] |
| | Detecting obstacles ahead | 19 (45.2%) | [31.7, 59.5] |
| | Route finding | 29 (69.1%) | [54.0, 80.9] |
| | Other challenges | 6 (14.3%) | [6.7, 27.6] |
| Obstacle Height | Head level | 32 (76.2%) | [61.5, 86.5] |
| | Chest level | 17 (40.5%) | [27.0, 55.5] |
| | Leg level | 16 (38.1%) | [25.0, 53.2] |
| | Ground-level | 17 (40.5%) | [27.0, 55.5] |
| | None | 1 (2.4%) | [0.4, 12.3] |

*Table 4 Technology nonuse and device preference by impairment severity: reported as n (%).*

| Severity | Nonuse | Wearable | Handheld | Other |
|---|---|---|---|---|
| Mild | 0 (0.0%) | 3 (75.0%) | 1 (25.0%) | 0 (0.0%) |
| Moderate | 2 (33.3%) | 5 (83.3%) | 0 (0.0%) | 1 (16.7%) |
| Severe | 3 (17.6%) | 10 (58.8%) | 6 (35.3%) | 1 (5.9%) |
| Total blindness | 5 (33.3%) | 8 (53.3%) | 5 (33.3%) | 2 (13.3%) |

Beyond device form, participants expressed preferences regarding how navigation information should be delivered. Audio and multimodal feedback were the preferred options, whereas tactile-only and simple auditory cues were selected less often, as shown in Table 5. Among tactile options, feedback delivered to the hands and arms was preferred most frequently.

*Table 5 Device and feedback preferences; reported as n (%), and 95% CI; multiple selections allowed for tactile feedback location.*

| | | | |
|---|---|---|---|
| Form | Wearable | 26 (61.9%) | [46.8, 75.0] |
| | Handheld | 12 (28.6%) | [17.2, 43.6] |
| | Other | 4 (9.5%) | [3.8, 22.1] |
| Feedback | Audio | 17 (40.5%) | [27.0, 55.5] |
| | Multimodal | 13 (31.0%) | [19.1, 46.0] |
| | Tactile | 8 (19.1%) | [10.0, 33.3] |
| | Beep | 4 (9.5%) | [3.8, 22.1] |
| Tactile Location | Hands and arms | 24 (57.1%) | [42.2, 70.9] |
| | Wrists | 16 (38.1%) | [25.0, 53.2] |
| | Shoulders | 12 (28.6%) | [17.2, 43.6] |
| | Forehead | 6 (14.3%) | [6.7, 27.8] |
| | Feet | 5 (11.9%) | [5.2, 25.0] |
| | Waist | 4 (9.5%) | [3.8, 22.1] |
| | Legs | 1 (2.4%) | [0.4, 12.3] |

Participants' expectations were also reflected in the route characteristics and system features they prioritized. Table 6 presents these route preferences and desired assistive navigation features. Obstacle-free paths and access to route information were selected most often, followed by continuous guidance. In addition, participants emphasized ease of use, low cost, reduced cognitive and physical effort, and effective obstacle avoidance. To facilitate comparison of the relative importance of system features, Figure 2 provides a ranked visualization of the responses summarized in Table 6.

*Table 6 Route preferences and desired assistive navigation features; reported as n (%), 95% CI; multiple selections allowed.*

| | | | |
|---|---|---|---|
| Route Factors | Shortest route | 16 (38.1%) | [25.0, 53.2] |
| | Fastest travel time | 15 (35.7%) | [23.0, 50.8] |
| | Lowest-cost path | 11 (26.2%) | [15.3, 41.1] |
| | Obstacle-free paths | 26 (61.9%) | [46.8, 75.0] |
| | Route information | 23 (54.8%) | [39.9, 68.8] |
| | Continuous guidance | 20 (47.6%) | [33.4, 62.3] |
| | Rapid rerouting | 16 (38.1%) | [25.0, 53.2] |
| Key Features | Low mental load | 22 (52.4%) | [37.7, 66.6] |
| | Low physical effort | 20 (47.6%) | [33.4, 62.3] |
| | Ease of use | 34 (81.0%) | [66.7, 90.0] |
| | Obstacle avoidance | 20 (47.6%) | [33.4, 62.3] |
| | Low device weight | 22 (52.4%) | [37.7, 66.6] |
| | Low cost | 27 (64.3%) | [49.2, 77.0] |
| | Remote assistance | 10 (23.8%) | [13.5, 38.5] |

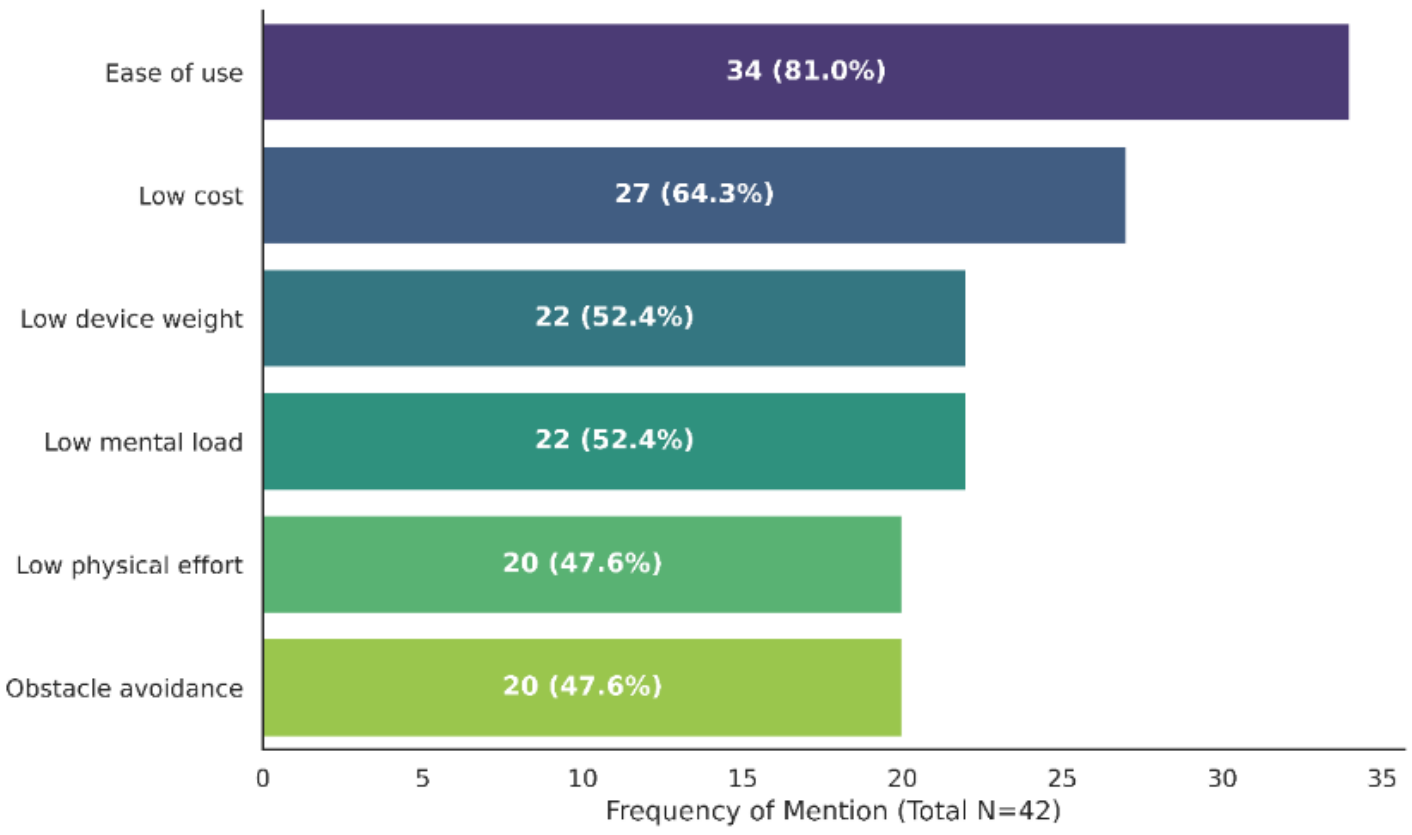

*Figure 2 Importance of assistive navigation system features based on participant responses.*

## Insights from Open-Ended Responses

To complement the quantitative findings, open-ended responses provide additional insights into user experiences and expectations. These responses emphasized simplicity, low-effort interaction, and integration with familiar, everyday devices. Participants preferred unobtrusive and discreet solutions that complement existing mobility aids, such as canes, rather than replacing them. Reliable and actionable guidance, including obstacle-related information and clear route instructions, was frequently requested. Practical considerations, particularly affordability, accessibility across regions, and operational constraints such as offline functionality, were also highlighted.

## Discussion

The findings of this study highlight persistent challenges faced by visually impaired individuals in wayfinding, spatial awareness, and obstacle avoidance. These challenges likely arise because real-world environments contain complex and multi-level obstacles that are not fully captured by current assistive technologies or traditional mobility aids. As a result, assistive navigation systems must improve their ability to detect and communicate spatial information across different obstacle levels, particularly beyond ground-level hazards. The results indicate that outdoor environments remain especially challenging, contradicting current research attention on indoor navigation. This can be explained by the higher variability and unpredictability of outdoor environments, as well as the lack of standardized infrastructure, where conditions such as noise, crowd density, and layout change continuously. Therefore, assistive navigation systems should prioritize robustness and adaptability to real-world conditions rather than relying on assumptions derived from structured or controlled environments.
A clear gap is observed between the availability of assistive navigation technologies and their actual use, with many participants reporting that they do not rely on digital tools. This pattern reflects practical barriers reported by participants, including limited access, high cost, and the difficulty of obtaining suitable technologies. It also highlights environmental constraints that may vary across countries, as well as reliance on alternative navigation strategies. These challenges are particularly relevant for users with higher levels of impairment. In addition, reports of insufficient feedback and limited responsiveness suggest that current systems may not deliver timely and actionable information in safety-critical situations.
A further discrepancy emerges between current technology use and user preferences. While most participants rely on smartphone-based navigation in practice, this appears to reflect availability rather than optimal usability. In contrast, a majority expressed a preference for wearable, hands-free solutions, indicating a shift toward interaction modalities that have the potential to reduce physical and cognitive burden. Together with the strong emphasis on ease of use and low cost, these findings indicate that nonuse is driven primarily by practical barriers rather than a lack of demand.
These findings further suggest that assistive navigation systems should support multiple device form factors rather than relying on a single solution. While wearable devices enable hands-free interaction, handheld solutions may still offer advantages in terms of control and familiarity, particularly for users with different levels of impairment and navigation strategies. Similarly, the diversity of preferred feedback modalities, including speech, multimodal, and tactile cues, reflects variability in environmental conditions and individual preferences. This indicates that no single feedback mechanism is universally effective, and that future systems should incorporate adaptable and multimodal interaction strategies that respond to both user needs and situational constraints.
Finally, participants prioritized obstacle-free routes, ease of use, and low cognitive effort over traditional efficiency metrics such as shortest path or travel time. This shift reflects the safety-critical nature of navigation for visually impaired users, where reliability and mental workload are more important than efficiency. As a result, evaluation frameworks for assistive navigation systems should move beyond purely technical metrics and incorporate user-centered criteria such as safety, cognitive load, and real-world usability.

## Limitations

This study has several limitations that should be considered when interpreting the results. First, the survey relied on online recruitment, which may introduce sampling bias toward individuals with internet access and familiarity with digital tools. As a result, individuals with limited access or those who rely on alternative communication methods such as Braille may be underrepresented. The survey was conducted exclusively in English, which excluded non-English speakers and limited representation across regions. Furthermore, the data collection period was constrained by time limitations, which may have influenced the overall sample size and distribution. Future studies addressing these limitations could contribute to a more comprehensive understanding of navigation challenges and needs among visually impaired individuals.

## Conclusion

This paper presents the results of a fully accessible online survey examining navigation experiences, challenges, and preferences among visually impaired individuals. The findings provide a user-centered view of everyday navigation needs and highlight the diversity of experiences across users and contexts. They further reveal a gap between current technology use and user preferences, with participants often relying on available handheld solutions while expressing a clear preference for wearable, hands-free systems. Participants emphasized ease of use, affordability, obstacle avoidance, and access to route information, indicating that assistive navigation is evaluated not only in terms of technical performance but also in terms of safety, cognitive effort, and practical usability. Therefore, future systems should support flexible device form factors and interaction modalities, prioritize safety and cognitive simplicity, and provide low-effort, adaptive, and multimodal feedback. Affordability and practical accessibility should be treated as central design requirements, while evaluation should extend beyond technical metrics to include usability, reliability, and real-world effectiveness. Despite limitations related to sampling and scope, this study contributes structured evidence that can inform future assistive navigation research and development.

## Acknowledgements

We would like to thank all survey participants for their time and valuable input. We are also grateful to Oriane Pierrès for reviewing the questionnaire and providing helpful feedback, and to Reyhaneh Fatemeh Bamdad for her contribution to the data analysis of this survey.

## References

1. Zimmermann-Janschitz, S., Mandl, B., Dückelmann, A.: Clustering the Mobility Needs of Persons with Visual Impairment or Legal Blindness. Transportation Research Record: Journal of the Transportation Research Board. 2650, 66–73 (2017).
2. Hara, K., Azenkot, S., Campbell, M., Bennett, C.L., Le, V., Pannella, S., Moore, R., Minckler, K., Ng, R.H., Froehlich, J.E.: Improving Public Transit Accessibility for Blind Riders by Crowdsourcing Bus Stop Landmark Locations with Google Street View: An Extended Analysis. ACM Transactions on Accessible Computing 6, 1–23 (2015). https://doi.org/10.1145/2717513.
3. Brunet, L., Darses, F., Auvray, M.: Strategies and needs of blind pedestrians during urban navigation. Le travail humain. 81, 141–171 (2018).
4. El-Taher, F.E.-Z., Miralles-Pechuán, L., Courtney, J., Millar, K., Smith, C., Mckeever, S.: A survey on outdoor navigation applications for people with visual impairments. IEEE Access. 11, 14647–14666 (2023).
5. Golledge, R.G., Marston, J.R., Costanzo, C.M.: Attitudes of Visually Impaired Persons toward the Use of Public Transportation. Journal of Visual Impairment & Blindness. 91, 446–459 (1997).
6. Parkin, J., Smithies, N.: Accounting for the Needs of Blind and Visually Impaired People in Public Realm Design. Journal of Urban Design. 17, 135–149 (2012).
7. Wong, S.: Traveling with blindness: A qualitative space-time approach to understanding visual impairment and urban mobility. Health & Place. 49, 85–92 (2018).
8. Jeamwatthanachai, W., Wald, M., Wills, G.: Indoor navigation by blind people: Behaviors and challenges in unfamiliar spaces and buildings. British Journal of Visual Impairment. 37, 140–153 (2019).
9. Papadopoulos, K., Charitakis, K., Koustriava, E., Kouroupetroglou, G., Stiefelhagen, R., Stylianidis, E., Gumus, S.S.: Environmental Information Required by Individuals with Visual Impairments Who Use Orientation and Mobility Aids to Navigate Campuses. Journal of Visual Impairment & Blindness. 114, 263–276 (2020).
10. Müller, K., Engel, C., Loitsch, C., Stiefelhagen, R., Weber, G.: Traveling More Independently: A Study on the Diverse Needs and Challenges of People with Visual or Mobility Impairments in Unfamiliar Indoor Environments. ACM Transactions on Accessible Computing. 15, 1–44 (2022). https://doi.org/10.1145/3514255.
11. Williams, M.A., Hurst, A., Kane, S.K.: “Pray before you step out”: describing personal and situational blind navigation behaviors. In: Proceedings of the 15th International ACM SIGACCESS Conference on Computers and Accessibility. pp. 1–8. ACM, Bellevue Washington (2013).
12. Chaudary, B., Pohjolainen, S., Aziz, S., Arhippainen, L., Pulli, P.: Teleguidance-based remote navigation assistance for visually impaired and blind people—usability and user experience. Virtual Reality. 27, 141–158 (2023).

13. Plikynas, D., Žvironas, A., Budrionis, A., Gudauskis, M.: Indoor navigation systems for visually impaired persons: Mapping the features of existing technologies to user needs. Sensors. 20, 636 (2020).
14. Barontini, F., Catalano, M.G., Pallottino, L., Leporini, B., Bianchi, M.: Integrating wearable haptics and obstacle avoidance for the visually impaired in indoor navigation: A user-centered approach. IEEE transactions on haptics. 14, 109–122 (2020).
15. Braun, V., Clarke, V.: Using thematic analysis in psychology. Qualitative Research in Psychology. 3, 77–101 (2006).
16. Agresti, A., Coull, B.A.: Approximate is Better than “Exact” for Interval Estimation of Binomial Proportions. The American Statistician. 52, 119–126 (1998).